\documentclass[twocolumn,showpacs,prb]{revtex4}

\usepackage{graphicx}
\usepackage{dcolumn}
\usepackage{bm}
\usepackage{times}

\begin{document}

\preprint{}

\title{Hole Transport in \textit{p}-Type ZnO}

\author{Takayuki~\textsc{Makino}}
 \affiliation{Department of Material Science, University of Hyogo, Ako-gun, Hyogo 678-1297 and The Institute of Physical and Chemical Research (RIKEN), Sendai 980-0845, Japan}
\author{Atsushi \textsc{Tsukazaki}}%

\author{Akira \textsc{Ohtomo}}
\author{M. \textsc{Kawasaki}}
\affiliation{%
Institute for Materials Research, Tohoku University,
Sendai 980-8577, Japan}%
\author{Hideomi \textsc{Koinuma}}
\affiliation{National Institute for Matrials Science, Tsukuba, Ibaraki 305-0044, Japan}%

\date{\today}

\begin{abstract}
A two-band model involving the A- and B-valence bands was adopted to
analyze the temperature dependent Hall effect measured on N-doped
\textit{p}-type ZnO. The hole transport characteristics (mobilities, and effective Hall factor) are calculated using the ``relaxation time approximation'' as a function of temperature. It is shown that the lattice scattering by the acoustic deformation potential is dominant. In the calculation of the scattering rate for ionized impurity mechanism, the activation energy of 100 or 170 meV is used at different compensation ratios between donor and acceptor concentrations. The theoretical Hall mobility at acceptor concentration of $7 \times 10^{18}$~cm$^3$ is about 70~cm$^2$V$^{-1}$s$^{-1}$ with the activation energy of 100~meV and the compensation ratio of 0.8 at 300~K. We also found that the compensation ratios
conspicuously affected the Hall mobilities.
\end{abstract}

\maketitle
\section{Introduction}

Recently, ZnO has received great attention due to its potential applications in ultraviolet (UV) and blue optoelectronic devices~\cite{look_sst,nature_mat_tsukazaki} and its unique material properties such as a large direct band gap of $\simeq$3.3~eV at room temperature, high transparency for visible light, mixability with Mg and Cd to form the quaternary system (Zn,Mg,Cd)O with a tunable band gap between 3.0 and 4.0~eV~\cite{makino14}. Nominally ZnO is grown under fairly residual \textit{n}-type. It is not difficult to achieve higher electron concentrations with excess zinc or by doping with Al, Ga, or In~\cite{makino19,makino_int_Ga,kim1,ellmer_mob1}.

On the other hand, it is necessary to obtain \textit{p}-type layers for light-emitter application and make a progress for its reliable technology. But, this has not been solved up to now and remains the major bottleneck for ZnO-based optoelectronics.

Because of the difficulties in growing \textit{p}-type ZnO films as earlier-mentioned, very little is known about the material and physical properties of this material. Fortunately, the situation is now getting improved and several groups have reported the growth of \textit{p}-type ZnO with valid and reasonable sets of experimental data. Among them, Look and coworkers~\cite{lookp-type} have reported on the temperature dependence of the Hall-effect measurement. Tsukazaki and coworkers have fabricated first ZnO-based light emitters~\cite{nature_mat_tsukazaki} in which they also reported a temperature-dependent Hall-effect (\textit{T}-Hall). In this work, we report mainly the theoretical aspects of the \textit{T}-Hall results and the comparison with the experimental data.

Several groups~\cite{rode_bkmob1,dclook_mob2,albrecht:6864,makino_trans} have already reported the electron transport characteristics of \textit{n}-ZnO both experimentally and theoretically. On one hand, there is few theoretical investigation~\cite{look_sst} of the hole transport characteristics of \textit{p}-ZnO. In this paper, the Hall coefficient anisotropy factors ($r_A$) for ZnO are determined and theoretical hole transport characteristics (Hall and drift mobilities, and effective Hall factor) are calculated. The calculated hole transport characteristics, with changing compensation ratio over a wide range of temperatures (\textit{T} = 50 to 400~K), assume transport in both A-valence-hole and B-valence-hole bands (a two-band model) with scattering allowed between these bands.

\section{Calculation Procedures}

The analysis follows essentially the same method used for representative
cubic semiconductors (GaAs, \textit{c}-GaN), but the calculation had to be modified to fit the valence band structures of wurtzitic ZnO. In addition, 
we take the effects of warping of surfaces into account for the hole transport studied here. The Hall coefficient factor, which generally links the experimentally
measured Hall mobility with the drift mobility can be written as follows:

\begin{equation}
r_i=\frac{<\tau_i^2>}{<\tau_i>^2},
\label{hall-fac1}
\end{equation}
where $\tau$ is the carrier scattering time, $<\quad >$ indicates a thermal average over the distribution of carrier energy and the index $i$ runs over A- ($i = 1$) and B-valence ($i = 2$) holes.

The Hall coefficient factor depends on the degree of warping at the warped
bands as well as the scattering mechanism, and can be written as:
\begin{equation}
r_i=r_{Ai} \frac{<\tau_i^2>}{<\tau_i>^2},
\label{hall-fac2}
\end{equation}
where $r_{Ai}$ is the anisotropy factor for \textit{i}-th hole band.

\begin{table}[htbp]
	\begin{center}
		\begin{tabular}{ccccccccc}
\hline
$A_1$&$A_2$&$A_3$&$A_4$&$A_5$&$A_6$&$-A$&$-B$&C$^2$\\
\hline
-6.680&-0.454&6.128&-2.703&-2.767&-4.626&1.41&1.95&0.796\\
\hline		
		\end{tabular}
	\end{center}
	\caption{Valence band parameters for ZnO. The notation of parameters used here are the same those in ref.~\onlinecite{hayashi1999}.}
	\label{tbl:1}
\end{table}

For their evaluation, one should determine the valence band parameters \textit{L}, \textit{M} and \textit{N} defined in terms of the interband matrix element. Then \textit{A}, \textit{B} and \textit{C} parameters (which are different from the notation related to the splitted valence bands of wurtzitic semiconductors), which have been used in an approximate expression for the B-valence- and A-valence-hole energy surfaces. We use here the cubic approximation. These valence band parameters were calculated from those of effective mass parameters ($A_1$ \textrm{to} $A_6$) recently reported by Fan and coworkers~\cite{wjfan2006}, which are compiled in Table~\ref{tbl:1}. The cubic approximation yielded the linkage of $A_1$ \textrm{to} $A_6$ with the parameters of \textit{L}, \textit{M} and \textit{N}, whose relations are given by:

\begin{equation}
L_1=A_2+A_4+A_5,
\label{luttinger1}
\end{equation}
\begin{equation}
L_2=A_1,
\end{equation}
\begin{equation}
M_1=A_2+A_4-A_5,
\end{equation}
\begin{equation}
M_2=A_1+A_3,
\end{equation}
\begin{equation}
M_3=A_2,
\end{equation}
\begin{equation}
N_1=2\times A_5,
\end{equation}
and 
\begin{equation}
N_2=\sqrt{2} \times A_6,
\end{equation}
where the notations introduced above are:
\begin{equation}
L=(L_1+L_2)\slash 2,
\label{luttinger7}
\end{equation}
\begin{equation}
M=(M_1+M_2+M_3)\slash 3,
\label{luttinger8}
\end{equation}
\begin{equation}
N=(N_1+N_2)\slash 2,
\label{luttinger9}
\end{equation}

The relationship between \textit{L}, \textit{M}, and \textit{N} and the well-known notations of \textit{A}, \textit{B} and \textit{C} has been given in ref.~\onlinecite{hayashi1999}.

Here, we used the formulae expressed in ref.~\onlinecite{wiley1} to calculate $r_{Ai}$, the detailed descriptions of which were given in eq.~(B.2) of ref.~\onlinecite{hayashi1999}.

The scattering mechanisms due to (1) deformation potential due to acoustic (dp) and optical phonons (npop), polar optical phonon (pop), piezoelectric (pe), ionized impurity (imp), and interband deformation potentials are taken into account
in the two-band analysis involving both kinds of hole scatterings. All scattering rates have been derived assuming a a nonparabolic band, which is given below, spherical bands, and band warping is accounted for.

The dispersion relation of energy band is represented for the nonparabolic
band picture in the following form:
\begin{equation}
\frac{\hbar^2 k^2}{2 m^*}=E (1+\xi E),
\label{kane1}
\end{equation}

where all the notations used take their conventional meanings. The detailed
explanation for the description can be found elsewhere~\cite{hayashi1999}.

The approximated expressions for the ``overlap function'' describing
the hole scattering from the initial state $\vec{k}$ in band $i (\vec{k}_i)$ to the final state $\vec{k}^\prime $ in band $f (k^\prime_f )$ are given by:
\begin{eqnarray}
G(\vec{k}_i,\vec{k}_f^{\prime}) = \left\{
	\begin{array}{@{\,}ll}
	(1+3 \cos^2 \theta)/4 & \textrm{for} \quad i=f \quad \textrm{intraband}\\
	3 (1-\cos^2 \theta)/4 & \textrm{for} \quad i\neq f \quad \textrm{interband}\\
	\end{array}
\right.
\label{oi}
\end{eqnarray}
where $\theta $ is the angle between the wavenumbers of these relevant two states.

We summarize the relevant equations related to the scattering mechanisms treated in this study. The screening effects for the phonon scattering mechanism can be neglected as far as the regime of low-impurity concentrations is concerned here.

The \textit{ionized impurity scattering} mechanism was treated using the Brooks-Herring approximation, where particles interact via the screened Coulombic potential~\cite{brooks1,herring1}. The inverse of screening length, $\lambda_D$, can be written for a low hole concentration, as follows:
\begin{equation}
\frac{1}{\lambda_D^2} =\frac{e^2}{\epsilon_s k_BT} \{ p+(p+N_D) \}[1-(N_D+p)/N_A] \}
\label{screen}
\end{equation}
where $p$ is the total hole concentration, and $\epsilon_s$ is the static dielectric constant. Other notations have their usual meanings. The free hole concentration $p$ can be calculated using equation (3.3) of ref.~\onlinecite{hayashi1999}.

The Hall-effect measurement evaluated the value of $E_a$ to be approximately 100~meV, while the spectroscopic studies experimentally yielded $E_a \simeq 170$~meV~\cite{lookp-type,bkmeyer-rev}. Therefore, $E_a$ used in this work is 100 or 170~meV.

The scattering rate due to the density of ionized impurities $N_I$ ($N_I = p + 2N_D$) and the level of ionization of the impurity atom $Z_I$ is given by:
\begin{equation}
S_{ion}(\vec{k}_i,\vec{k}_f^{\prime})=\frac{Z_I^2 N_I e^4 m^*_f}{8 \pi \epsilon_\infty \hbar^3}\frac{1}{k_i^3} \int_{-1}^{1} \frac{(1-y) G(y)}{(1-y+1/2k_i^2\lambda_D^2)^2} dy,
\label{scat-ion}
\end{equation}
where $y \equiv \cos \theta $ and $G(y) = G(\cos \theta) $ is the overlap function in eq.~(\ref{oi}). Also, $m^*_f$ is the effective mass of the relevant band into which the holes are being scattered, $k_i$ is the magnitude of the initial-state wave vector and the other notations take their conventional meanings.

The scattering rate for the \textit{nonpolar optical scattering} has been given elsewhere~\cite{hayashi1999}. Quantity of $D_{npo}$ is found to be about 3.9$\times 10^{11}$~eV/m. We list here the scattering rate for \textit{polar optical mechanism } because of some possible typographical error appearing in ref.~\onlinecite{hayashi1999}:
\begin{eqnarray}
S_{po} (\vec{k}_i, \vec{k}_f^{\prime})=\frac{e^2 m_f^* \omega_0}{4 \pi \hbar^2} 
\left( \frac{1}{\epsilon_\infty}- \frac{1}{\epsilon_s} \right) 
\times \frac{1}{k_i^2} \times \cr
\left( \sum_{+,-} k_{\pm} \left( n_0 +\frac{1}{2} \mp \frac{1}{2} \right) 
\times \int_{-1}^{1} \frac{G(y)}{[1-2 (k_{\pm}/k_i) +(k_{\pm}/k_i)^2]} dy \right).
\label{scat-po}
\end{eqnarray}
where $\epsilon_\infty$ is the high-frequency dielectric constant. The wavevectors $k_{\pm}$ correspond to energies $E_i \pm \hbar \omega_0$. The equations in regard with the scatterings by \textit{acoustic deformation potential } and by the \textit{piezo-scattering} have been enumerated in ref.~\onlinecite{hayashi1999} as equations (c.2) and (c.3).

In order to determine drift and Hall mobilities from the above-mentioned scattering mechanisms, we have determined the effective Hall factor using the ``relaxation time approximation'' or otherwise known as ``Mathiessen's rule''.
The effective Hall factor $r_{\rm eff}$ is:
\begin{equation}
r_{\rm eff}=\frac{(1+\alpha^{3/2})(r_1 \alpha^{3/2} + \beta^2 r_2)}{(1+\alpha^{3/2} \beta)^2},
\label{effec-Hall-fac}
\end{equation}
with, $\alpha = m_1=m_2$, $\beta=\mu_1/\mu_2$, $\mu_i = e <\tau_i> /m^{*}_i$, and $r_i = r_{Ai} <\tau_i^2>/<\tau_i>^2$. Quantities $m_i$ $\mu_i$, and $<\tau_i>$ are the effective masses, drift mobilities and the relaxation times for the \textit{i}-th hole band, respectively. (The index $i = 1$ always refers to A-valence, and the index $i = 2$ to B-valence holes.)

The drift mobility $\mu$ is determined from relationships between drift and Hall mobilities given by:
\begin{equation}
\mu=\frac{\mu_H}{r_{\rm eff}},
\label{drift}
\end{equation}
where $\mu_H$ is the Hall mobility and $r_{\rm eff}$ is the effective Hall factor given by eq.~(\ref{effec-Hall-fac}).

The Hall mobility, $\mu_H$, is determined to be:
\begin{equation}
\mu_H=\frac{e}{m^*_1} \frac{r_{A1} <\tau_1^2> +\alpha^{1/2}r_{A2}<\tau_2^2>}{<\tau_1>+\alpha^{-1/2} <\tau_2>},
\label{Hall-mob}
\end{equation}
where $m^*_1$ is the effective A-valence-hole mass, and $r_{A1}$ and $r_{A2}$ are given in eq.~(B.2) of ref.~\onlinecite{hayashi1999}.

\begin{table}[htbp]
	\begin{center}
		\begin{tabular}{cccc}
Property& &Units&Value\\
\hline
$E_g$&Energy gap for A-valence hole&eV&3.445\\
$E_g$&Energy gap for B-valence hole&eV&3.455\\
$m_1/m_0$&Effective A-valence hole mass ratio& &0.59\\
$m_2/m_0$&Effective B-valence hole mass ratio& &0.59\\
$\rho$&Density&kg/m$^3$&5.67526\\
$s$&Sound velocity&m/s&6.09$\times 10^5$\\
$E_1$&Acoustic deformation potential&eV&15.0\\
$D_{npo}$&Optical coupling constant&eV/m&3.9$\times 10^{11}$\\
$\epsilon_s $&Static dielectric constant& &8.2\\
$\epsilon_\infty $&High-frequency dielectric constant& &3.7\\
$\hbar \omega_0$&Polar optical phonon energy&eV&0.072\\
$h_{pz}$&Piezoelectric constant&C/m$^2$&0.89\\
\hline
		\end{tabular}
	\end{center}
	\caption{Material parameters of ZnO used in the calculation.}
	\label{tbl:2}
\end{table}

The material parameters used in the calculations on ZnO are listed in Table~\ref{tbl:2}. Rode~\cite{rode_bkmob1} suggests a theoretical value of 3.8~eV for deformation potential $E_1$, but the later studies resulted in a better agreement
with the experimental data if the larger value was used.

\section{Results and Discussion}

We determined the valence band parameters (\textit{L, M, N, A, B, C}) using Luttinger parameters. Their results are listed in Table~I. The evaluated
anisotropy factors $r_{A1}$ and $r_{A2} $ both yielded the values very
close to the unity, which is significantly different from the situation observed in GaN. This difference may be due to the isotropic nature of the hole masses in ZnO~\cite{LBZincoxide}.

\begin{figure}[htbp]
	\includegraphics[width=0.45\textwidth]{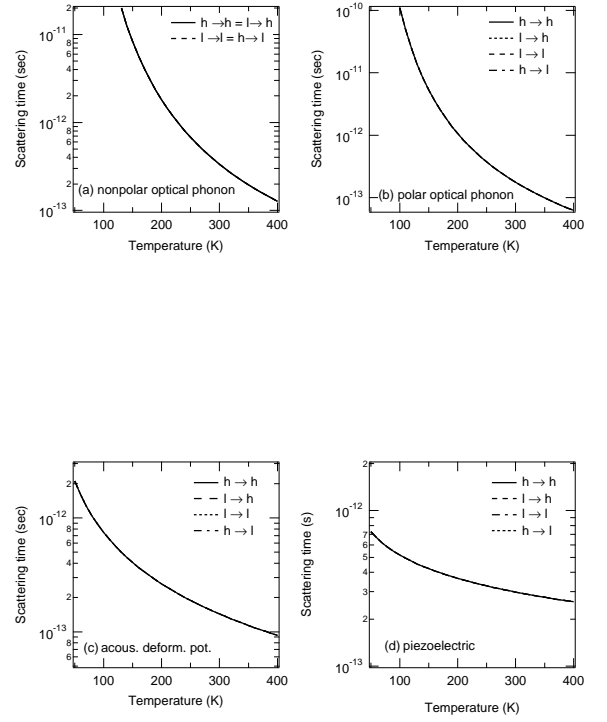}
	\caption{Variation with temperature of calculated scattering relaxation time corresponding to nonpolar optical phonon mode (a), polar optical phonon mode (b), acoustic deformation potential mode (c) and piezoelectric mode (d). Many traces are almost overlapped each other because the two hole masses are the same in ZnO. (h$\to $h - scattering from A-valence-to-A-valence-hole band, h$\to $l - scattering from A-valence-to-B-valence-hole band, l$\to $l - scattering from B-valence-to-B-valence-hole band, and
l$\to $h- scattering from B-valence-to-A-valence-hole band.)}
	\label{scat-lattice}
\end{figure}

The partial `$\mu$ versus temperature' curves are plotted in Fig.~\ref{scat-lattice}. The interband scattering characteristic due to nonpolar optical phonons shown in Fig. \ref{scat-lattice}(a) is different from that due to other scattering mechanisms shown in Figs. \ref{scat-lattice}(b), \ref{scat-lattice}(c) and \ref{scat-lattice}(d). This is because the scattering relaxation time due to nonpolar optical phonon scattering is determined by the final-state effective hole mass ($m^*_f$), while that due to other scattering mechanisms is determined by both $m^*_f$ and the magnitude of the initial state of the wave vector ($k_i$) in which nonparabolicity is considered. We remark that, the
relative importance of the various scattering mechanisms mimics
that found in ZnSe~\cite{ruda_mob1}. For example, at temperatures below 300~K, the acoustic deformation potential scattering plays the most important role. 

\begin{figure}[htbp]
	\includegraphics[width=0.45\textwidth]{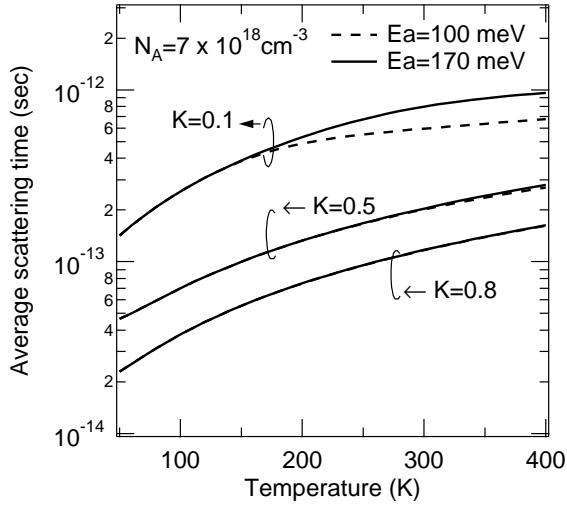}
	\caption{Variation of calculated average scattering time for ionized impurity scattering with temperature for $E_a = 100$ and 170~meV with dependence
on compensation ratio ($K/N_D=N_A$). The acceptor concentration ($N_A$)
was set to 7 $\times  10^{18}$~cm$^{-3}$ in the calculations.}
	\label{scat-ion}
\end{figure}

Fig.~\ref{scat-ion} plotted average scattering time ($\tau_{av}$) by ionized impurities calculated for two activation energies ($E_a = 100$ and 170~meV) with compensation ratios ($K = N_D/N_A$) between 0.1 and 0.8. Value of $\tau_{av}$ is calculated with $\tau_{scat}^{-1}$ as the sum of the scattering rate from A-to A-hole, A-to-B-hole, B-to-B-hole and B to- A-hole bands.

In the range of those acceptor activation energies ($E_a = 100-170$~meV), the effect of $E_a$ on the average scattering time at high temperatures is fairly negligible ($\tau_{av}(170)/\tau_{av}(100) \simeq 1$ at 300~K), while the corresponding effect for a low compensation ratio (\textit{K} = 0.1) is appreciable ($\tau_{av}(170)/\tau_{av}(100) \simeq 2$ at 300~K).

\begin{figure}[htbp]
	\includegraphics[width=0.45\textwidth]{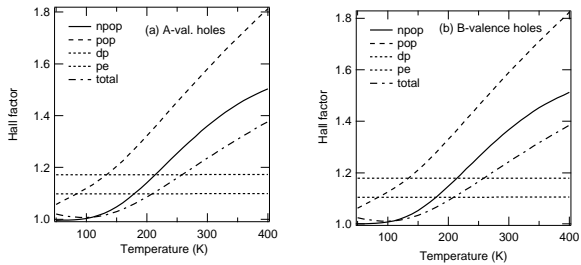}
	\caption{Variation of calculated effective Hall factors with temperature due to lattice scattering mechanisms for A-valence holes (a) and for B-valence holes (b). \textbf{npop} is the nonpolar optical phonon scattering. \textbf{pop} is the polar optical phonon scattering. \textbf{dp} is the acoustic phonon scattering. \textbf{pe} is the piezoelectric scattering. \textbf{total} is the total of all four lattice scattering mechanisms.}
	\label{hall-fac-lattice}
\end{figure}

The temperature dependences of Hall factors for the four different scattering mechanisms plotted in Fig.~\ref{hall-fac-lattice}. The Hall factors for A-valence-hole and B-valence-hole bands of nonpolar and polar optical scattering vary with temperature, while those of the acoustic deformation potential and piezoelectric scattering remained nearly unchanged, the different observations of which may be attributed to the temperature dependence of the scattering relaxation time of respective participating phonons. The values of the Hall factor both for A-valence-hole bands and for B-valence-hole bands varied from 1.0 to 1.5.

\begin{figure}[htbp]
	\includegraphics[width=0.3\textwidth]{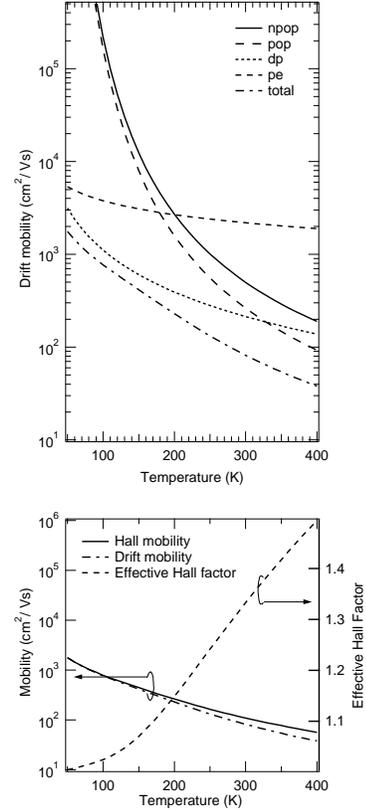}
	\caption{Theoretical drift mobilities due to each lattice scattering mechanisms (a), and the theoretical Hall and drift mobilities and the effective Hall factor (b) due to lattice scattering mechanisms as a function of temperature.}
	\label{mobility-lattice}
\end{figure}

The partial Hall and drift mobilities, and the effective Hall factor that we calculated are plotted in Fig.~\ref{mobility-lattice} against temperature. The acoustic phonon scattering is the most important mechanism, limiting the hole mobility over a wide range of temperature. For temperatures above 300 K, the polar optical phonon scattering mechanism is the dominant factor.

The obtained effective Hall factors vary from 1.1 to 1.7 on changing the temperature, as shown in Fig.~\ref{mobility-lattice}(b), indicating the importance of proper considerations of the Hall factors at high temperatures when comparing the drift mobility with $\mu_H$.

\begin{figure}[htbp]
	\includegraphics[width=0.45\textwidth]{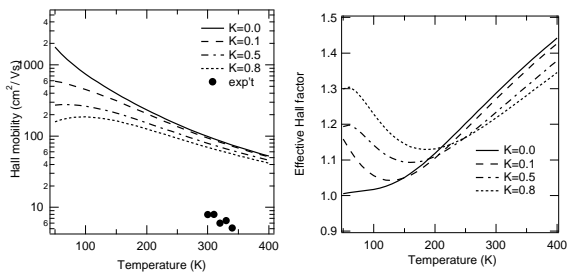}
	\caption{Theoretical Hall mobilities (a), and effective Hall factor (b) due to lattice and ionized impurity scattering mechanisms for different ratio of compensation as a function of temperature. $E_a$ used in the ionized impurity scattering mechanism is 100~meV. Also shown in (a) is a set of experimental data of ZnO:N thin film for comparison.}
	\label{mobility-total}
\end{figure}
Finally, Fig.~\ref{mobility-total} corresponds to the similar plot with Fig.~\ref{mobility-lattice}, including now the ionized impurity scattering with four different compensation ratios (\textit{K}) between 0.0 and 0.8. The acceptor concentration is $N_A = 7 \times 10^{17}$ cm$^{-3}$ and $E_a = 100$~meV. We also plotted the experimental data of \textit{p}-ZnO with nitrogen concentration of $2\times 10^{20}$~cm$^{-3}$ (different from the acceptor concentration $N_A$) for comparison. The compensation ratio and free hole concentration at 330~K of this sample are $p \approx 8 \times 10^{16}$~cm$^{-3} $ and approximately $K=0.8$~\cite{nature_mat_tsukazaki}.

Unfortunately, the calculated Hall mobility is still higher than that in the experimental reports. We considered only dp, npop, pop, and imp scattering
mechanisms in our modeling, which may be a result of this poor agreement.
A dislocation scattering and the space-charge effect probably also contribute
as a scattering mechanism of holes to the mobility determination.
Actually, it is known that the former mechanism have an influence on
relatively-poor-quality \textit{n}-type ZnO epitaxial layers~\cite{makino-pss-trans}.
We would like to propose more experiments using further optimized samples be done in future.

With an increaase in the compensation ratio (\textit{K}), the peaking temperature at maximum mobility also increases. The effective Hall factor ($r_{\rm eff}$) increases as \textit{K} decreases with increasing temperature. From Figs.~\ref{mobility-total}(a) and \ref{mobility-total}(b), it is observed that the maximum value of the Hall and drift mobilities are in the temperature range 50 to 100~K for all values of \textit{K} between 0.1 and 0.8. In comparison with Fig. \ref{mobility-lattice}(b), Fig.~\ref{mobility-total}(b) shows that the inclusion of scattering by ionized impurities into the Hall factor calculation is important at high compensation ratios (\textit{K} = 0.1 to 0.8), exhibiting wavy behavior with the change in temperature. The obtained effective Hall factors vary from 1.1 to 1.5 with changing temperature ($K \ge 0.1$). The temperature at which the effective Hall factor is minimized and the effective Hall factor is affected by the ionized impurity scattering mechanism.

\section{Conclusions}

In this study, a two-band model analysis was carried out in detail
on nitrogen-doped ZnO epitaxial layers. The theoretical Hall and drift mobilities, and the effective Hall factor were calculated at temperatures of 50 to 400~K. The hole transport is characterized by the scattering by the acoustic deformation potentials at temperatures below 330~K. The theoretical Hall mobility at acceptor concentration $N_A = 7 \times 10^{18}$~cm$^{-3}$ is about 70~cm$^2$/V$^{-1}$s$^{-1}$ with the activation energy of 100~meV and the compensation ratio of 0.8 at 300~K. The compensation ratio strongly affected the hole transport
characteristics.

\begin{acknowledgments}
We wish to thank Y. Segawa at the Institute of Physical and Chemical Research
(RIKEN) and Y. Takagi at the University of Hyogo for helpful discussions. This project was supported by Iketani Foundation of Science and Technology, Japan
under Contract No. 0181027A.
\end{acknowledgments}

\begin{thebibliography}{10}

\bibitem{look_sst}
D.~C. Look, Semicond. Sci. \& Technol. {\bf 20},  S55  (2005).

\bibitem{nature_mat_tsukazaki}
A. Tsukazaki, A. Ohtomo, T. Onuma, M. Ohtani, T. Makino, M. Sumiya, K. Ohtani,
  S.~F. Chichibu, S. Fuke, Y. Segawa, H. Ohno, H. Koinuma, and M. Kawasaki,
  Nat. Mater. {\bf 4},  42  (2005).

\bibitem{makino14}
T. Makino, Y. Segawa, M. Kawasaki, A. Ohtomo, R. Shiroki, K. Tamura, T. Yasuda,
  and H. Koinuma, Appl. Phys. Lett. {\bf 78},  1237  (2001).

\bibitem{makino19}
T. Makino, K. Tamura, C.~H. Chia, Y. Segawa, M. Kawasaki, A. Ohtomo, and H.
  Koinuma, Phys. Rev. B {\bf 65},  121201(R)  (2002).

\bibitem{makino_int_Ga}
T. Makino, Y. Segawa, A. Tsukazaki, A. Ohtomo, S. Yoshida, and M. Kawasaki,
  Appl. Phys. Lett. {\bf 85},  759  (2004).

\bibitem{kim1}
K.~J. Kim and Y.~R. Park, Appl. Phys. Lett. {\bf 78},  475  (2001).

\bibitem{ellmer_mob1}
K. Ellmeer, J. Phys. D. {\bf 34},  3097  (2001).

\bibitem{lookp-type}
D.~C. Look, D.~C. Reynolds, C.~W. Litton, R.~L. Jones, D.~B. Eason, and G.
  Cantwell, Appl. Phys. Lett. {\bf 81},  1830  (2002).

\bibitem{rode_bkmob1}
D.~L. Rode, Semiconductors and Semimetals {\bf 10},  1  (1975).

\bibitem{dclook_mob2}
D.~C. Look, D.~C. Reynolds, J.~R. Sizelove, R.~L. Jones, C.~W. Litton, G.
  Cantwell, and W.~C. Harsch, Solid State Commun. {\bf 105},  399  (1998).

\bibitem{albrecht:6864}
J.~D. Albrecht, P.~P. Ruden, S. Limpijumnong, W.~R.~L. Lambrecht, and K.~F.
  Brennan, Journal of Applied Physics {\bf 86},  6864  (1999).

\bibitem{makino_trans}
T. Makino, Y. Segawa, A. Tsukazaki, A. Ohtomo, and M. Kawasaki, Appl. Phys.
  Lett. {\bf 87},  022101  (2005).

\bibitem{wjfan2006}
W. Fan, J. Xia, P.~A. Agus, S. Tan, S. Yu, and X. Sun, J. Appl. Phys. {\bf 99},
   013702  (2006).

\bibitem{hayashi1999}
Y. Hayashi, K. Watanabe, T. Jimbo, and M. Umeno, Jpn. J. Appl. Phys. {\bf 37},
  622  (1999).

\bibitem{wiley1}
J.~D. Wiley, {\em Semiconductors and Semimetals}, {\em Vol. 10} (Academic, New
  York, 1975), chap. 2.

\bibitem{brooks1}
G.~B. Benedek, W. Paul, and H. Brooks, Phys. Rev. {\bf 100},  1129  (1955).

\bibitem{herring1}
C. Herring, Bell System Tech. J. {\bf 34},  1  .

\bibitem{bkmeyer-rev}
B.~K. Meyer, H. Alves, D.~M. Hofmann, W. Kriegseis, D. Forster, F. Bertram, J.
  Christen, A. Hoffmann, and M. Strassburg, Phys. Status Solidi (B) {\bf 241},
  231  (2004).

\bibitem{LBZincoxide}
E. Mollwo,  in {\em Semiconductors: Physics of II-VI and I-VII Compounds,
  Semimagnetic Semiconductors}, Vol.~17 of {\em Landolt-B{\"{o}}rnstein New
  Series}, edited by O. Madelung, M. Schulz, and H. Weiss (Springer, Berlin,
  1982), p.\ 35.

\bibitem{ruda_mob1}
H.~E. Ruda and B. Lai, J. Appl. Phys. {\bf 68},  1714  (1990).

\bibitem{makino-pss-trans}
T. Makino, Y. Segawa, A. Tsukazaki, A. Ohtomo, and M. Kawasaki, Phys. Status
  Solidi (c) {\bf 3},  956  (2006).

\end{thebibliography}


\end{document}